\documentclass[superscriptaddress,showpacs,aps,prd]{revtex4}
\pdfoutput=1
\usepackage{graphicx}
\usepackage{multibib}
\newcommand{\epp}{$e^+e^- \to \eta\pi^+\pi^-$ }

\begin{document}
\title{\large \bf \boldmath 
Measurement of the  $e^+e^- \to \eta\pi^+\pi^-$ cross section
in the center-of-mass energy range 1.22--2.00 GeV with the SND detector 
at the VEPP-2000 collider}

\author{V. M. Aulchenko}
\affiliation{Budker Institute of Nuclear Physics, SB RAS,
Novosibirsk, 630090, Russia}
\author{M.~N.~Achasov}
\affiliation{Budker Institute of Nuclear Physics, SB RAS,
Novosibirsk, 630090, Russia}
\affiliation{Novosibirsk State University, Novosibirsk, 630090, Russia}
\author{A.~Yu.~Barnyakov}
\affiliation{Budker Institute of Nuclear Physics, SB RAS,
Novosibirsk, 630090, Russia}
\author{K.~I.~Beloborodov}
\author{A.~V.~Berdyugin}
\author{A. G. Bogdanchikov}
\affiliation{Budker Institute of Nuclear Physics, SB RAS,
Novosibirsk, 630090, Russia}
\affiliation{Novosibirsk State University, Novosibirsk, 630090, Russia}
\author{A.~A.~Botov}
\affiliation{Budker Institute of Nuclear Physics, SB RAS,
Novosibirsk, 630090, Russia}
\author{T.~V.~Dimova}
\author{V.~P.~Druzhinin}
\author{V.~B.~Golubev}
\author{L.~V.~Kardapoltsev}
\affiliation{Budker Institute of Nuclear Physics, SB RAS,
Novosibirsk, 630090, Russia}
\affiliation{Novosibirsk State University, Novosibirsk, 630090, Russia}
\author{A.~G.~Kharlamov}
\affiliation{Budker Institute of Nuclear Physics, SB RAS,
Novosibirsk, 630090, Russia}
\affiliation{Novosibirsk State University, Novosibirsk, 630090, Russia}
\author{I.~A.~Koop}
\author{A.~A.~Korol}
\affiliation{Budker Institute of Nuclear Physics, SB RAS,
Novosibirsk, 630090, Russia}
\affiliation{Novosibirsk State University, Novosibirsk, 630090, Russia}
\author{S.~V.~Koshuba}
\author{D.~P.~Kovrizhin}
\affiliation{Budker Institute of Nuclear Physics, SB RAS,
Novosibirsk, 630090, Russia}
\author{A.~S.~Kupich}
\affiliation{Budker Institute of Nuclear Physics, SB RAS,
Novosibirsk, 630090, Russia}
\affiliation{Novosibirsk State University, Novosibirsk, 630090, Russia}
\author{K.~A.~Martin}
\author{A.~E.~Obrazovsky}
\affiliation{Budker Institute of Nuclear Physics, SB RAS,
Novosibirsk, 630090, Russia}
\author{E.~V.~Pakhtusova}
\affiliation{Budker Institute of Nuclear Physics, SB RAS,
Novosibirsk, 630090, Russia}
\affiliation{Novosibirsk State University, Novosibirsk, 630090, Russia}
\author{A.~I.~Senchenko}
\affiliation{Budker Institute of Nuclear Physics, SB RAS,
Novosibirsk, 630090, Russia}
\author{S.~I.~Serednyakov}
\author{Z.~K.~Silagadze}
\author{Yu.~M.~Shatunov}
\affiliation{Budker Institute of Nuclear Physics, SB RAS,
Novosibirsk, 630090, Russia}
\author{P. Ju. Shatunov}
\affiliation{Budker Institute of Nuclear Physics, SB RAS,
Novosibirsk, 630090, Russia}
\affiliation{Novosibirsk State University, Novosibirsk, 630090, Russia}
\author{D.~A.~Shtol}
\email[e-mail:]{D.A.Shtol@inp.nsk.su}
\affiliation{Budker Institute of Nuclear Physics, SB RAS, Novosibirsk, 630090,
Russia}
\author{D.~B.~Shwartz}
\affiliation{Budker Institute of Nuclear Physics, SB RAS,
Novosibirsk, 630090, Russia}
\affiliation{Novosibirsk State University, Novosibirsk, 630090, Russia}
\author{A.~N.~Skrinsky}
\affiliation{Budker Institute of Nuclear Physics, SB RAS,
Novosibirsk, 630090, Russia}
\author{I.~K.~Surin}
\affiliation{Budker Institute of Nuclear Physics, SB RAS,
Novosibirsk, 630090, Russia}
\author{Yu.~A.~Tikhonov}
\affiliation{Budker Institute of Nuclear Physics, SB RAS,
Novosibirsk, 630090, Russia}
\affiliation{Novosibirsk State University, Novosibirsk, 630090, Russia}
\author{Yu.~V.~Usov}
\affiliation{Budker Institute of Nuclear Physics, SB RAS,
Novosibirsk, 630090, Russia}
\author{A.~V.~Vasiljev}
\affiliation{Budker Institute of Nuclear Physics, SB RAS,
Novosibirsk, 630090, Russia}
\affiliation{Novosibirsk State University, Novosibirsk, 630090, Russia}

\begin{abstract}
In the experiment with the SND detector at the VEPP-2000 $e^+e^-$ collider
the cross section for the process \epp has been measured in the center-of-mass
energy range from 1.22 to 2.00 GeV. Obtained results are in agreement 
with previous measurements and have better  accuracy.
The energy dependence of the \epp cross section has been fitted with
the vector-meson dominance model. From this fit the product of the 
branching fractions $B(\rho(1450)\to\eta\pi^+\pi^-)B(\rho(1450)\to e^+e^-)$ 
has been extracted and compared with the same products for 
$\rho(1450)\to\omega\pi^0$ and $\rho(1450)\to\pi^+\pi^-$ decays.
The obtained cross section data have been also used to test the conservation
of vector current hypothesis.
\end{abstract}

\maketitle

\section{Introduction}
\begin{figure}
\includegraphics[width=0.60\textwidth]{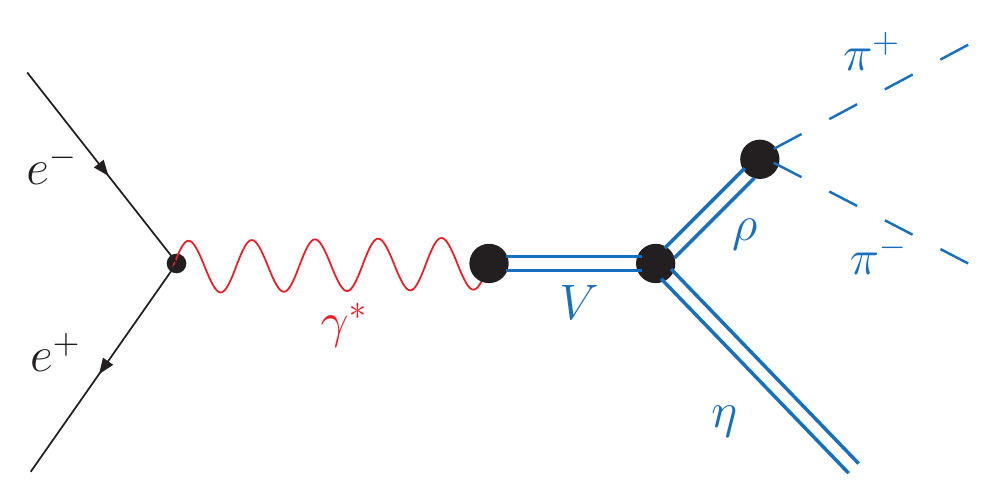}
\caption{The diagram for the process $e^+e^- \to \eta \pi^+\pi^-$ 
in the VMD model.\label{fig:diagram}}
\end{figure}
The process \epp contributes to the 
isovector part of the electromagnetic hadronic current.
In the vector-meson dominance (VMD) model it is described by the 
diagram shown in Fig.~\ref{fig:diagram}, where $V$ are
$\rho(770)$, $\rho(1450)$, and $\rho(1700)$ resonances.
In contrast to the main isovector 
modes $e^+e^-\to \pi^+\pi^-\pi^0\pi^0$ and $\pi^+\pi^-\pi^+\pi^-$ the
process \epp is dominated by one intermediate state only, $\eta\rho(770)$,
and therefore important for determination of $\rho(1450)$ and $\rho(1700)$ 
resonance parameters.
The process gives a sizable contribution, up to 5\% at center-of-mass
(c.m.) energy $\sqrt{s}=1.5$ GeV, to 
the total hadronic cross section, which is determined below 2 GeV as
a sum of exclusive modes. 
Data on \epp can be used to test the conservation of vector current
(CVC) hypothesis, which predicts a relation between the 
\epp cross section and the spectral function for the 
$\tau^- \to \eta \pi^-\pi^0\nu_{\tau}$ decay.

The \epp process was earlier studied in several
experiments~\cite{bib:nd,bib:dm2,bib:cmd2,bib:babar,bib:eppvepp2m}.
The most precise measurements were performed at the VEPP-2M $e^+e^-$ collider
with the CMD-2~\cite{bib:cmd2} and SND~\cite{bib:eppvepp2m} detectors
below 1.4 GeV, and at PEP-II B-factory with the BABAR detector~\cite{bib:babar}
above 1.4 GeV using the initial state radiation technique.
In the CMD-2 and BABAR measurements the $\eta$ meson was reconstructed
via its decay to $\pi^+\pi^-\pi^0$, while SND used the 
$\eta\to \gamma \gamma$ decay mode.
This work continues the SND study of Ref.~\cite{bib:eppvepp2m} in a wider
energy region, up to 2 GeV, using data collected at the 
VEPP-2000 $e^+e^-$ collider~\cite{bib:vep2000}.

\section{Experiment}
SND is a nonmagnetic detector consisting of a tracking system, 
aerogel threshold Cherenkov counters for kaon identification,
an electromagnetic calorimeter, and a muon system. 
The tracking system based on a nine-layer drift chamber provides
solid angle coverage of 94\% of 4$\pi$ and azimuthal and polar angle 
resolutions of $0.45^\circ$ and $0.8^\circ $, respectively. 
The three-layer spherical electromagnetic calorimeter contains 1640 NaI(Tl) 
crystals with a total thickness of 13.4$X_0$, where $X_0$ is the radiation 
length. A solid angle covered by the calorimeter is 90\% of 4$\pi$. Its energy 
resolution for photons is 
$\sigma_{E_\gamma}/E_\gamma = 4.2\%/\sqrt[4]{E_\gamma({\rm GeV})}$, and 
the angular resolution about $1.5^\circ$.

The experiment was performed at the VEPP-2000 in 2011--2012.
The c.m. energy range $\sqrt{s}$ = 1.05--2.00 GeV was
scanned several times with a step of 25 MeV. The total integrated 
luminosity collected by SND in this energy range is about 35 pb$^{-1}$. 
The analysis was performed initially for 2011 and 2012 data separately.
Since the cross sections measured in the two data sets are found to be
consistent, data collected at close energies in 2011 and 2012 are combined
in the analysis presented in this paper.

\section{Event selection}
\label{sec:sel}
Preliminary selection of \epp ($\eta \to \gamma\gamma$) event candidates is
based on the following requirements:
\begin{itemize}
  \item $N_c=2$, where $N_c$ is the number of charged particles originating
    from the interaction region. Each charged-particle track must 
    cross at least four drift-chamber layers and has $r_i<0.3$ cm and 
    $|Z_i|<10$ cm, where $r_i$ is the distance between the track and the beam axis,
    and $Z_i$ is the $z$-coordinate
    of the track at its distance of the closest approach to the beam axis.
  \item $N_\gamma$=2, where $N_\gamma$ is the number of reconstructed photons.
    The photon polar angle must be in the range 
    $36^\circ<\theta_{\gamma}<144^\circ$.
  \item $0.4<E_{\mathrm{tot}}/\sqrt{s}<0.9$ and $E_{\mathrm{char}}/\sqrt{s}<0.65$, 
    where $E_{\mathrm{tot}}$ is the total energy deposition in the calorimeter,
    and $E_{\mathrm{char}}$ is the total energy deposition in the calorimeter from
    charged particles. These conditions suppress QED background.
\end{itemize}

For selected events we perform a geometrical fit to a common vertex 
and a two-constrained kinematic fit to the $e^+e^- \to \pi^+\pi^-\gamma\gamma$
hypothesis, and then apply the following additional conditions:
\begin{itemize}
  \item $\chi^2_{\mathrm{vertex}}<200$, where $\chi^2_{\mathrm{vertex}}$ is $\chi^2$ of the 
    vertex fit.
  \item $\chi^2_{\pi^+\pi^-\gamma\gamma}<60$, where 
    $\chi^2_{\pi^+\pi^-\gamma\gamma}$ is $\chi^2$
    of the kinematic fit.
  \item $400~\textrm{MeV}\leq m_{\gamma\gamma}\leq 700~\textrm{MeV}$, where 
    $m_{\gamma\gamma}$ is the two-photon invariant mass calculated using
    photon parameters after the kinematic fit.
\end{itemize}

\section{Background subtraction}
\label{sec:subbg}
\begin{figure*}
    \begin{minipage}[t]{0.48\textwidth}
	\includegraphics[width=0.99\textwidth]{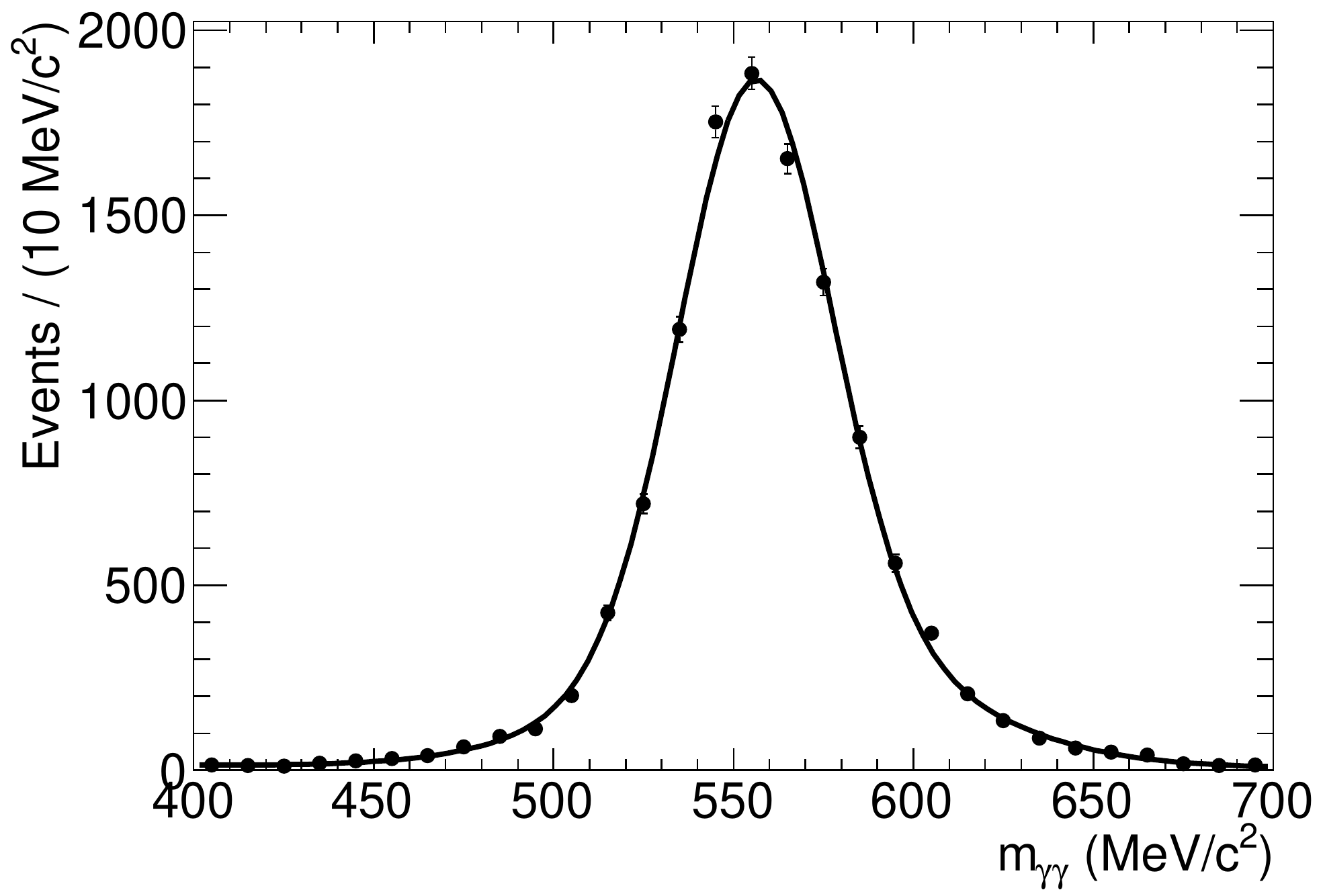}
	\caption{The two-photon invariant-mass spectrum for simulated \epp events at
	  $\sqrt{s}=1.5$~GeV (points with error bars) fitted with the double-Gaussian function.
	  \label{fig:imggmc}}
    \end{minipage}
    \hfill
    \begin{minipage}[t]{0.48\textwidth}
	\includegraphics[width=0.99\textwidth]{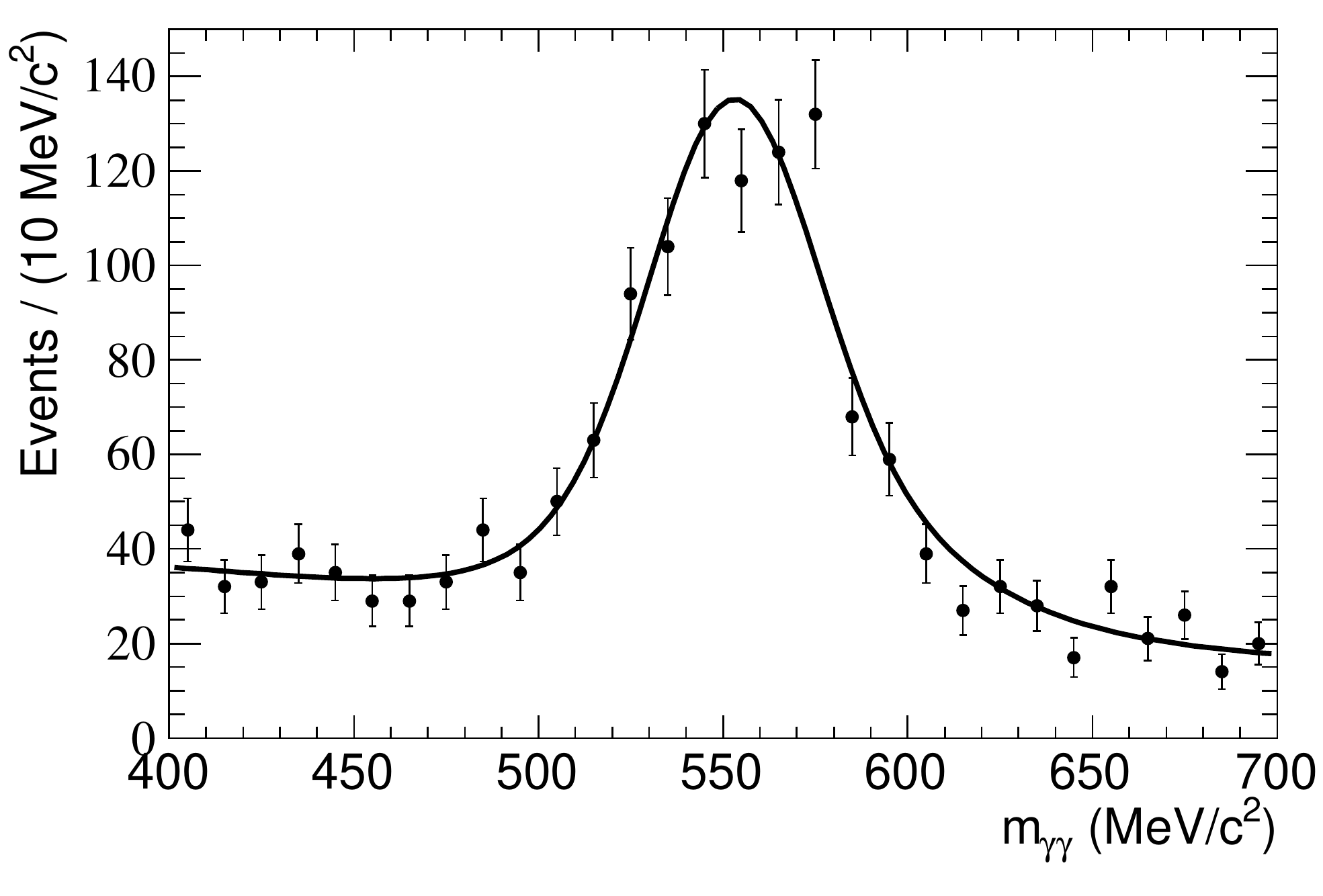}
	\caption{The two-photon invariant-mass spectrum for data events selected in the 
	  energy range $\sqrt{s}=1.45-1.60$~GeV. The curve is the result of the fit 
	  described in the text.
	  \label{fig:imggascorr}}
    \end{minipage}
\end{figure*}
Main background sources for the process under study are
the QED process $e^+e^- \to e^+e^-\gamma\gamma$, and multipion 
processes, e.g. $e^+e^- \to \pi^+\pi^-\pi^0\pi^0$. Events of
these processes are strongly suppressed by our selection criteria and
do not have a peak at the $\eta$ mass in the two-photon invariant mass
spectrum. 

The only source of peaking background, the process $e^+e^- \to \eta K^+K^-$,
is suppressed by the condition on $\chi^2_{\pi^+\pi^-\gamma\gamma}$.
Its contribution estimated using Monte Carlo (MC) simulation
and the $e^+e^- \to \eta K^+K^-$ cross section measured in 
Ref.~\cite{bib:etakk} is found to be less than 0.15\% and neglected.

To separate signal and background we fit to the two-photon invariant mass
spectrum with a sum of signal and background distributions.The signal 
line shape is described by a double-Gaussian function, parameters of
which are determined from a fit to the two-photon mass spectrum
for simulated \epp events. An example of such a fit at $\sqrt{s}=1.5$ GeV
is shown in Fig.~\ref{fig:imggmc}.

To take into account a possible difference between data and simulation
in the $\eta$ peak position and two-photon mass resolution, we introduce two
additional parameters, the mass shift $\Delta M$ 
($m_{1,2}=m_{1,2}^{\rm MC}+\Delta M$) and a 
width correction $\Delta\sigma^2$
($\sigma_{1,2}^2=(\sigma_{1,2}^{\rm MC})^2+\Delta\sigma^2$), where $m_{1,2}^{\rm MC}$ and 
$\sigma_{1,2}^{\rm MC}$
are the means and $\sigma$'s of the double-Gaussian function determined 
from simulation, and
$m_{1,2}$ and $\sigma_{1,2}$ are the corrected values of these parameters. 

The parameters $\Delta\sigma^2$ and $\Delta M$ are determined from the fit to the
spectrum for data events from the energy interval near the maximum of the
\epp cross section ($\sqrt{s}=1.45-1.60$ GeV). The spectrum and fitted
curve are shown in Fig.~\ref{fig:imggascorr}. The nonpeaking background
is described by a linear function. This assumption about the 
background shape was 
tested on simulated events of the dominant background process 
$e^+e^- \to \pi^+\pi^-\pi^0\pi^0$.
The found values of correction parameters
($\Delta M=-3.0\pm0.9$ MeV/$c^2$ and $\Delta\sigma^2=-89\pm33$ MeV$^2/c^4$ 
for the 2011 data set and
$\Delta M=-1.5\pm1.5$ MeV/$c^2$ and $\Delta\sigma^2=104\pm65$ MeV$^2/c^4$
for the 2012 data set)
and the assumption of linear background are used 
in the fits to data spectra for individual energy points. 
The difference between the correction parameters
for 2011 and 2012 is due to difference in angular
resolutions for charged tracks which is not
taken into account in simulation. 

The numbers of fitted \epp events for different energy points are listed in 
Table~\ref{tab:bcrs}. We do not observe any excess of signal events over 
background at energies below 1.22 GeV.

The data mass spectra in the three energy regions, 
1.20--1.45 GeV, 1.45--1.60 GeV, and 1.60--2.00 GeV, are also fit with
a quadratic background. The difference between the fits with the two
background hypotheses is taken as an estimate of the systematic
uncertainty due to the unknown background shape. It is found to be
6.7\% below 1.45 GeV, 1.0\% in the energy range 1.45--1.60 GeV,
and 2.2\% above.

\section{\boldmath Internal structure of the $\eta\pi^+\pi^-$ final state}
\label{sec:etarho}
\begin{figure*}
    \begin{minipage}{0.48\textwidth}
	\includegraphics[width=0.99\textwidth]{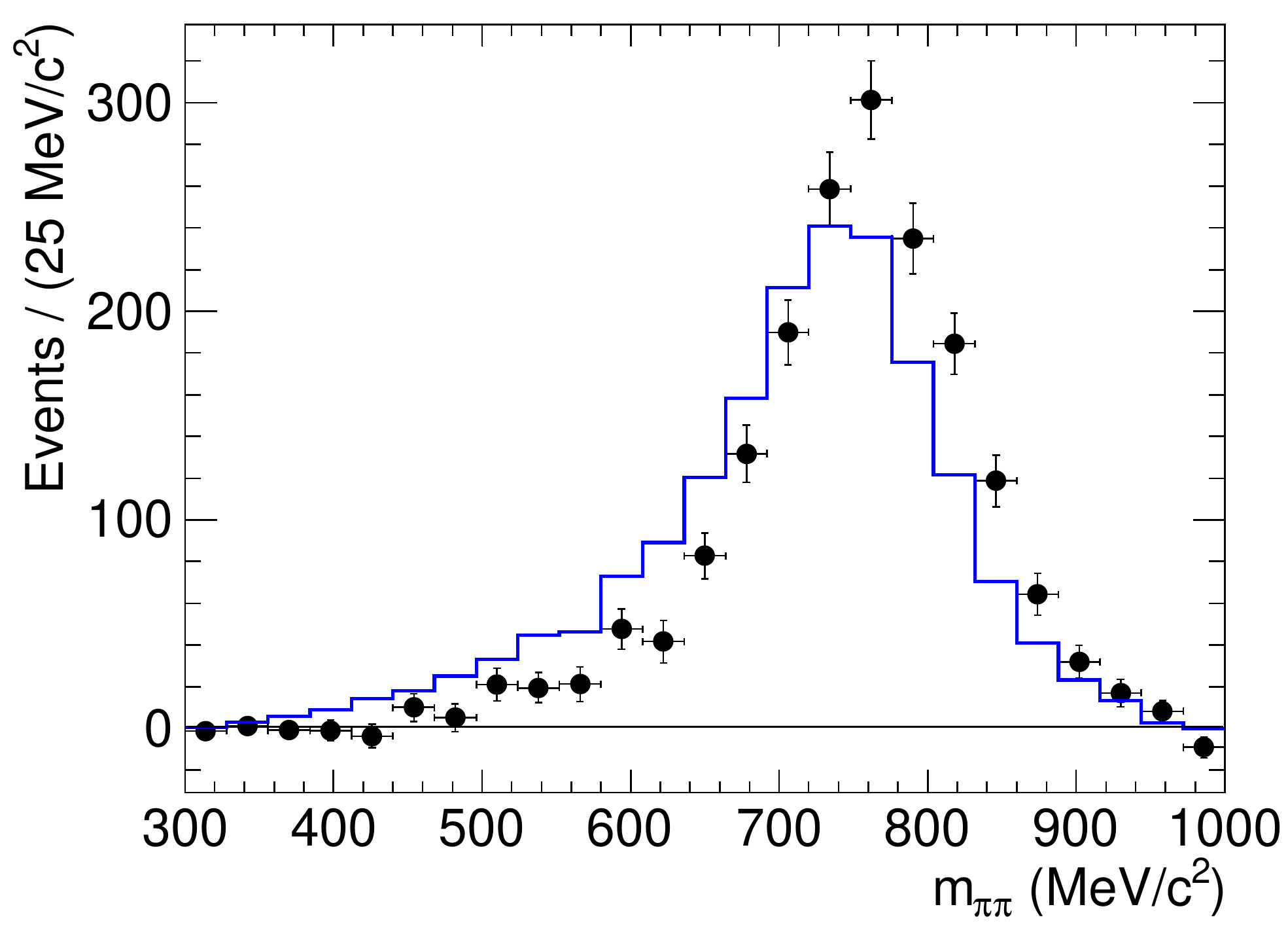}
	\caption{The $\pi^+\pi^-$ invariant-mass spectrum
	  for data (points with error bars) and simulated (histogram) \epp events 
	  from the energy range $\sqrt{s}=1.45-1.60$~GeV. The simulation uses
	  a model of the $\eta\rho(770)$ intermediate state.
	  \label{fig:pipi}}
    \end{minipage}
    \hfill
    \begin{minipage}{0.48\textwidth}
	\includegraphics[width=0.99\textwidth]{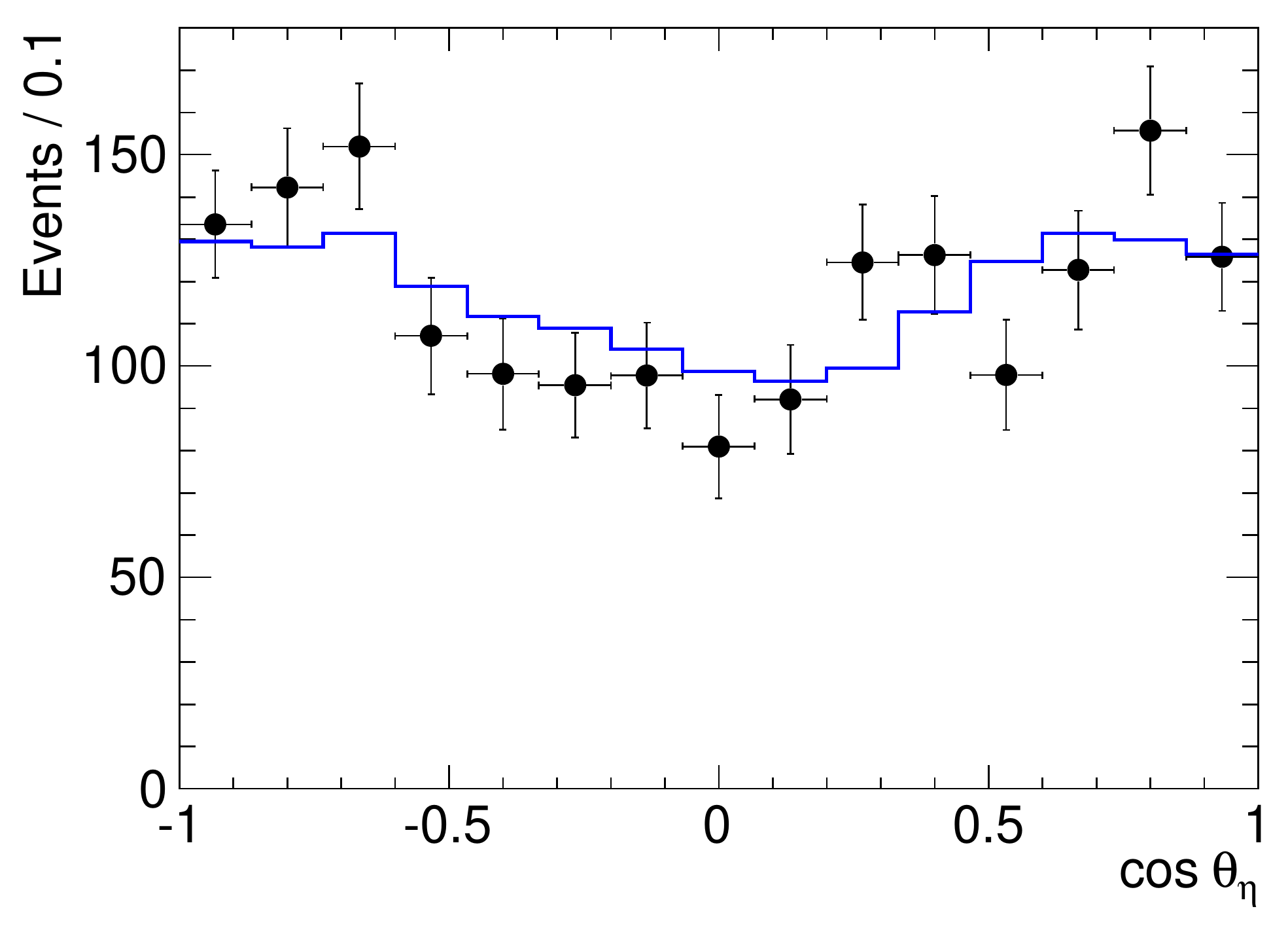}
	\caption{The $\cos{\theta_\eta}$ distribution for data (points with error bars)
	  and simulated (histogram) \epp events from the energy range 
	  $\sqrt{s}=1.45-1.60$~GeV. The simulation uses
	  a model of the $\eta\rho(770)$ intermediate state.
	  \label{fig:etatheta}}
    \end{minipage}
\end{figure*}
The $\pi^+\pi^-$ invariant mass ($m_{\pi\pi}$) spectrum for \epp data events
from the energy range $\sqrt{s}=$1.45--1.60 GeV is shown in
Fig.~\ref{fig:pipi}. The spectrum is obtained as a difference of the 
$\pi^+\pi^-$ mass spectrum for events with $500<m_{\gamma\gamma}<600$ MeV/$c^2$
and the spectrum for events from the sidebands 
($400<m_{\gamma\gamma}<470$ MeV/$c^2$ and 
$630 <m_{\gamma\gamma}<700$ MeV/$c^2$) divided by a scale 
factor of 1.4. The solid histogram in Fig.~\ref{fig:pipi}
is the result of MC simulation. The \epp event generator is based on formulas 
for the differential cross section from Ref.~\cite{bib:thepp} and uses
the model of the $\eta\rho(770)$ intermediate state. 
The observed difference between data and MC spectra is too large to be 
explained by imperfect simulation of resolution effects, and may be
a result of the contribution of other intermediate state, e.g. 
$\eta\rho(1450)$, and its interference with the dominant $\eta\rho(770)$
amplitude. A similar effect was observed, for example, in the $J/\psi \to 3\pi$ 
decay~\cite{bib:jpsi3pi}, in which the Dalitz plot distribution deviates
from the prediction for the $\rho\pi$ intermediate state.

In Fig.~\ref{fig:etatheta} we compare the data and simulated 
$\cos{\theta_\eta}$ distributions, where $\theta_\eta$ is 
the $\eta$-meson polar angle. In the $\eta\rho$ model this 
distribution is expected to be $1+\cos^2\theta_\eta$. We see reasonable
agreement between data and simulation in the angular distributions.

\section{Detection efficiency}
\label{sec:eff}
\begin{figure}
    \includegraphics[width=0.60\textwidth]{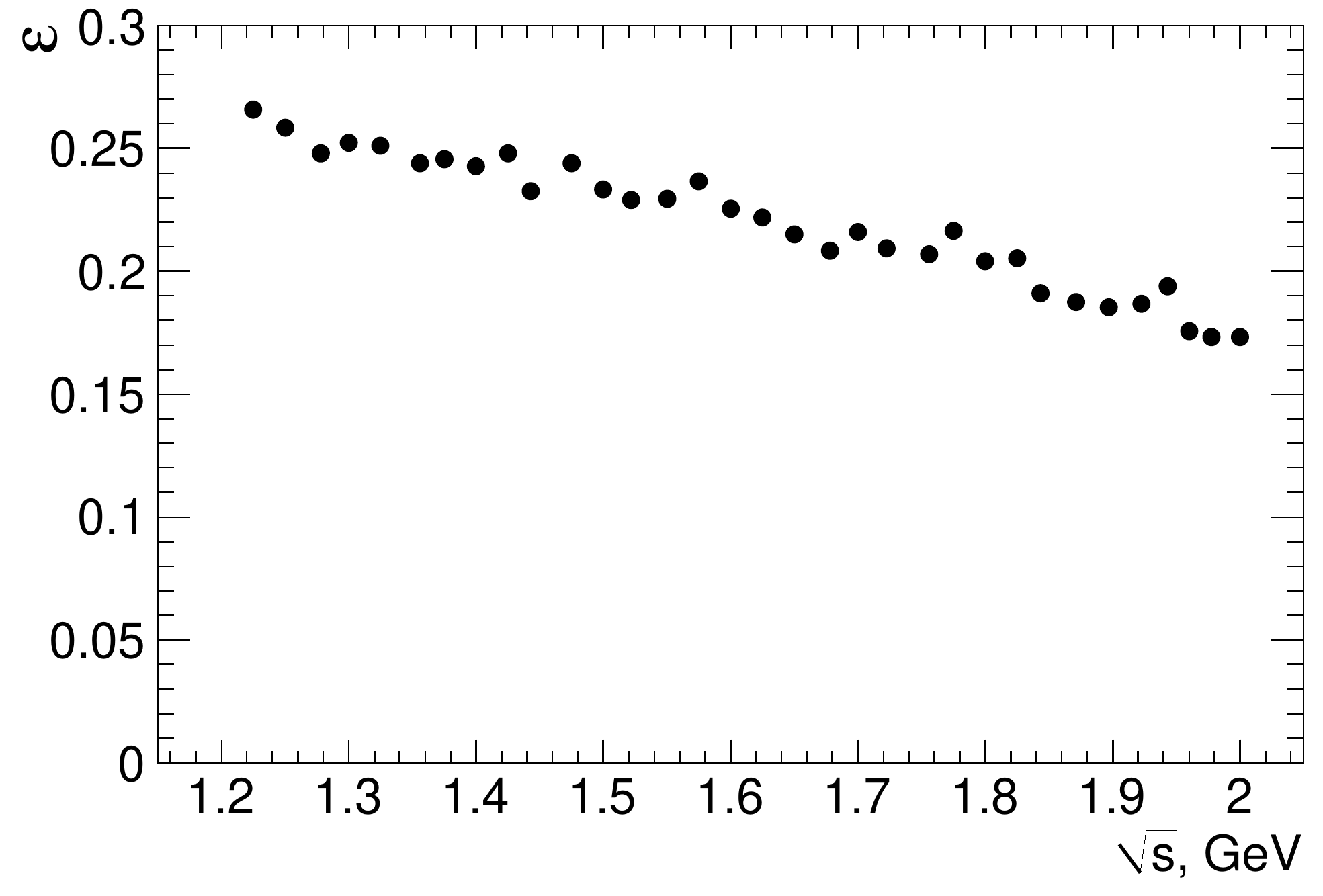}
    \caption{The detection efficiency for simulated 
      $e^+e^- \to \eta\pi^+\pi^-$, $\eta \to \gamma\gamma$ events.
      \label{fig:eff}}
\end{figure}
The detection efficiency for the process under study is determined using
MC simulation in the $\rho(770)\eta$ model. To estimate an influence of
the deviation from this model observed in the previous section, we reweight 
simulation events according to the $m_{\pi\pi}$ spectrum observed in data. 
The shift in the detection efficiency, about 1\%, is taken as an estimate 
of the model uncertainty associated with the $\rho(770)\eta$ assumption.

The simulation takes into account radiative corrections to the Born cross 
section calculated according to Ref.~\cite{bib:radcor}. In particular, an 
extra photon emitted by initial particles is generated with the angular 
distribution modeled according to Ref.~\cite{bib:BM}. Such an approach requires
knowledge of the energy dependence of the \epp cross section. This dependence
was taken initially from Ref.~\cite{bib:babar}. Then we repeat the simulation
with the energy dependence measured in this work. The variation of the
detection efficiency, less than 1.0\% at $\sqrt{s}<1.6$ GeV and less than 
4.2\% at higher energies, is considered as an estimate of the model error. 
The total model uncertainty of the detection efficiency is 1.4\% at 
$\sqrt{s}<1.6$ GeV and 4.3\% at $\sqrt{s}>1.6$ GeV.

Beam background overlapping with signal events can produce additional
clusters in the calorimeter and tracks in the tracking system. 
To take into account this effect in MC simulation, beam-background events 
recorded during experiment with a special random trigger are merged
with simulated events. The presence of beam-generated tracks and 
clusters in the calorimeter reduces detection efficiency,
by about 10\%. 

The energy dependence of the detection efficiency is shown in 
Fig.~\ref{fig:eff}. Nonmonotonic behavior of the efficiency as a 
function of energy is due to variations of experimental conditions 
(beam background, dead detector channels, etc.), which are taken into account
in MC simulation.

The detection efficiency obtained using MC simulation is corrected
to take into account a difference between data and simulation in detector 
response: $\varepsilon=\varepsilon_{\rm MC}(1-\Delta)$. To determine $\Delta$,
events from the energy region $\sqrt{s}=$1.45--1.60 GeV are used. We loosen 
a selection criterion, fit to $m_{\gamma\gamma}$ spectrum, and study variation 
in the fitted number of \epp events in data and simulation. The efficiency
correction for the tested criterion is determined from the data-MC simulation
double ratio $\Delta=(N^\ast/N)_{\rm data}/(N^\ast/N)_{\rm MC}-1$, 
where $N$ and $N^\ast$ are the fitted numbers of \epp events selected with 
standard and loosened criteria. 

To study the effect of the condition $N_\gamma=2$, we perform a kinematic 
fit for events with more than two photons. From all possible two-photon 
combinations in an event we choose the combination with 
$400~\textrm{MeV}\leq m_{\gamma\gamma}\leq 700~\textrm{MeV}$ and 
minimum $\chi^2_{\pi^+\pi^-\gamma\gamma}$.
From the fit to the $m_{\gamma\gamma}$ spectrum for these events
we determine $(N^\ast-N)$, and calculate the efficiency correction.
To determine correction for the condition $N_c=2$, we study events
with $N_c=3$. In the kinematic fit, two tracks with minimal $r_i$ are
used. For other selection criteria we shift the boundaries of the conditions,
from 60 to 10000 for $\chi^2_{\pi^+\pi^-\gamma\gamma}$, from 200 to 10000
for $\chi^2_{vtx}$, etc.

The resulting corrections are summarized in Table~\ref{tab:syst}. Listed are
those conditions for which statistically significant deviations of the
data-MC double ratios are observed. The two last rows in Table~\ref{tab:syst}
represent corrections for the data-MC difference in the ratio of the 
charged-track reconstruction efficiency for pions and 
electrons~\cite{bib:pipi}, and in the probability of photon conversion in the 
material before the drift chamber~\cite{bib:kardapo}.
\begin{table}
\caption{The efficiency corrections
\label{tab:syst}}
\begin{ruledtabular}
\begin{tabular}{lc}
 & $\Delta$, \% \\
\hline
Condition $N_\gamma=2$ & $-7.5 \pm 2.3$  \\
Condition $N_c=2$      &  $2.0 \pm 0.7$  \\
Condition $\chi^2_{\pi^+\pi^-\gamma\gamma}<60$ &  $-4.8  \pm 3.0$ \\
Condition $\chi^2_{\mathrm{vertex}}<200$     &  $0.9  \pm 0.4$ \\
Conditions $E_{\mathrm{tot}}<0.9\sqrt{s}$ and $E_{\mathrm{char}}<0.65\sqrt{s}$ & $-1.3 \pm 0.6$ \\
Track reconstruction & $0.3 \pm 0.2$ \\
Photon conversion     &  $0.4 \pm  0.6$ \\
\hline
Total                 &  $-10.1\pm3.9$
\end{tabular}
\end{ruledtabular}
\end{table}

The corrected values of the detection efficiency are listed in 
Table~\ref{tab:bcrs}. The statistical error on the detection efficiency is 
about 1\%. So, the total uncertainty on the detection efficiency including the
statistical error, the uncertainty in the efficiency correction, and the model
uncertainty is 4.3\% at $\sqrt{s}<1.6$ GeV and 6.0\% at $\sqrt{s}>1.6$ GeV.

\section{Luminosity measurement}
\label{sec:ilum}
Integrated luminosity is determined using large-angle Bhabha scattering
($e^+e^- \to e^+e^-$) events selected with the following criteria:
\begin{itemize}
  \item $N_c=2$ (see the $N_c$ definition in Sec.~\ref{sec:sel});
  \item $50^\circ<\theta_{1,2}<130^\circ$, where $\theta_{1,2}$ 
    are the polar angles of the charged particles;
  \item $E_{1,2}/\sqrt{s} > 0.25$, $0.65<(E_1+E_2)/\sqrt{s}<1.1$, 
    where $E_1$ and $E_2$ are the energies of the charged particles measured 
    in the calorimeter;
  \item $|\Delta\theta|<20^\circ$, $|\Delta\phi|<5^\circ$, where $\Delta\theta$
    and $\Delta\phi$ are the polar and azimuthal acollinearity angles.
\end{itemize}

To calculate the detection efficiency and the cross section for
the large-angle Bhabha scattering, the BHWIDE~\cite{bib:BHWIDE} event
generator is used. The integrated luminosity measured for each energy 
point ($L_i$) is listed in Table~\ref{tab:bcrs}.
The theoretical uncertainty on the cross section calculation
is better than 0.5\%. The systematic uncertainty on the detection
efficiency is estimated to be 2\%.
\begin{table}
\caption{
The c.m. energy ($\sqrt{s}$), integrated luminosity ($L$), detection efficiency ($\varepsilon$),
number of selected signal events ($N$ ), radiative-correction factor ($1 + \delta$), 
measured $e^+e^- \to \eta \pi^+\pi^-$ Born cross section ($\sigma_B$). 
For the number of events and cross section the statistical error is quoted.
The systematic uncertainty on the cross section is 8.3\% at $\sqrt{s}<1.45$ GeV, 
5.0\% at $1.45<\sqrt{s}<1.60$ GeV, and 7.8\% at $\sqrt{s}>1.60$ GeV.
\label{tab:bcrs}}
\begin{ruledtabular}
\begin{tabular}{cccccc}
$\sqrt{s}$, GeV & $\sigma_B$, nb & $N$ & $\varepsilon$ & $L$, nb$^{-1}$ & $1+\delta$   \\
\hline
1.225 & $0.35 \pm 0.15$ & $20 \pm 9 $ & 0.105  & 553 & 0.87 \\
1.250 & $0.17 \pm 0.15$ & $8  \pm 7 $ & 0.102  & 466 & 0.87 \\
1.278 & $0.49 \pm 0.13$ & $56 \pm 16$ & 0.097  &1225 & 0.87 \\
1.300 & $0.50 \pm 0.19$ & $23 \pm 10$ & 0.099  & 484 & 0.87 \\
1.325 & $0.74 \pm 0.19$ & $38 \pm 10$ & 0.099  & 542 & 0.86 \\
1.356 & $1.07 \pm 0.15$ & $137\pm 21$ & 0.096  &1398 & 0.86 \\
1.375 & $1.25 \pm 0.22$ & $70 \pm 13$ &  0.097 & 599 & 0.86 \\
1.400 & $1.69 \pm 0.24$ & $100\pm 15$ &  0.095 & 643 & 0.87 \\
1.425 & $2.23 \pm 0.25$ & $125\pm 15$ &  0.097 & 591 & 0.87 \\
1.443 & $2.76 \pm 0.19$ & $355\pm 26$ &  0.091 &1442 & 0.88 \\
1.475 & $3.31 \pm 0.29$ & $191\pm 18$ &  0.096 & 608 & 0.89 \\
1.500 & $3.63 \pm 0.29$ & $244\pm 20$ &  0.092 & 731 & 0.90 \\
1.522 & $4.47 \pm 0.23$ & $568\pm 30$ &  0.090 &1395 & 0.91 \\
1.550 & $4.28 \pm 0.33$ & $225\pm 18$ &  0.090 & 566 & 0.93 \\
1.575 & $3.61 \pm 0.36$ & $154\pm 16$ &  0.093 & 436 & 0.94 \\
1.600 & $3.30 \pm 0.34$ & $139\pm 15$ &  0.089 & 446 & 0.96 \\
1.625 & $3.76 \pm 0.34$ & $189\pm 17$ &  0.087 & 530 & 0.98 \\
1.650 & $2.53 \pm 0.32$ & $116\pm 15$ &  0.085 & 490 & 0.99 \\
1.678 & $2.41 \pm 0.19$ & $290\pm 23$ &  0.082 &1314 & 1.01 \\
1.700 & $2.79 \pm 0.31$ & $126\pm 14$ &  0.085 & 472 & 1.01 \\
1.723 & $2.05 \pm 0.20$ & $193\pm 19$ &  0.082 &1022 & 1.01 \\
1.756 & $2.26 \pm 0.19$ & $249\pm 20$ &  0.081 &1198 & 1.02 \\
1.775 & $1.97 \pm 0.28$ & $91 \pm 13$ &  0.085 & 473 & 1.03 \\
1.800 & $2.09 \pm 0.17$ & $274\pm 22$ &  0.080 &1391 & 1.06 \\
1.825 & $1.47 \pm 0.24$ & $74 \pm 12$ &  0.081 & 513 & 1.09 \\
1.843 & $1.36 \pm 0.15$ & $173\pm 18$ &  0.075 &1369 & 1.11 \\
1.871 & $0.94 \pm 0.13$ & $137\pm 18$ &  0.074 &1555 & 1.14 \\
1.897 & $0.89 \pm 0.11$ & $171\pm 20$ &  0.073 &2033 & 1.17 \\
1.922 & $0.81 \pm 0.13$ & $100\pm 15$ &  0.073 &1256 & 1.20 \\
1.943 & $0.75 \pm 0.12$ & $102\pm 15$ &  0.076 &1312 & 1.22 \\
1.960 & $0.76 \pm 0.17$ & $52 \pm 11$ &  0.069 & 724 & 1.24 \\
1.978 & $0.81 \pm 0.15$ & $86 \pm 14$ &  0.068 &1125 & 1.25 \\
2.000 & $0.84 \pm 0.21$ & $47 \pm 10$ &  0.068 & 576 & 1.28 \\
\end{tabular}	 			    				    
\end{ruledtabular}
\end{table}

\section{Results and discussion}
The Born cross section at the $i$th energy point 
is determined as:
\begin{equation}
    \sigma_\mathrm{B}^i=\frac{N_i}{\varepsilon_i L_i (1+\delta_i)},
    \label{eq:expbcrs}
\end{equation}
where $\delta_i$ is the radiative correction. 
Knowledge of the \epp Born cross section at energies below $\sqrt{s_i}$ 
is required to calculate $\delta_i$:
\begin{equation}
    1+\delta_i=\frac{\sigma_{\mathrm{vis}}(s_i)}{\sigma_{\mathrm{B}}(s_i)},
\end{equation}
\begin{equation}
    \sigma_{\mathrm{vis}}(s)=
    \int\limits_0^1\sigma_\mathrm{B}(s(1-z))F(z,s)dz,
    \label{eq:radcor}
\end{equation}
where $F(z,s)$ is a function describing the probability to emit extra photons
with the total energy $z\sqrt{s}/2$~\cite{bib:radcor}. Technically, the 
radiative corrections are calculated using the VMD model for the Born cross 
section described below. Parameters of the model are determined from a fit 
with Eq.~(\ref{eq:radcor}) to the measured visible cross 
section $N_i/(\varepsilon_i L_i)$. The obtained values of the 
radiative correction are listed in Table~\ref{tab:bcrs}. The model uncertainty
on the radiative correction is estimated by variation of the model parameters 
within their errors and is found to be 0.6\% below 1.45 GeV, 1.4\% in the 
energy range 1.45--1.60 GeV, and 4.1\% above 2.00 GeV.

\begin{figure}
\includegraphics[width=0.60\textwidth]{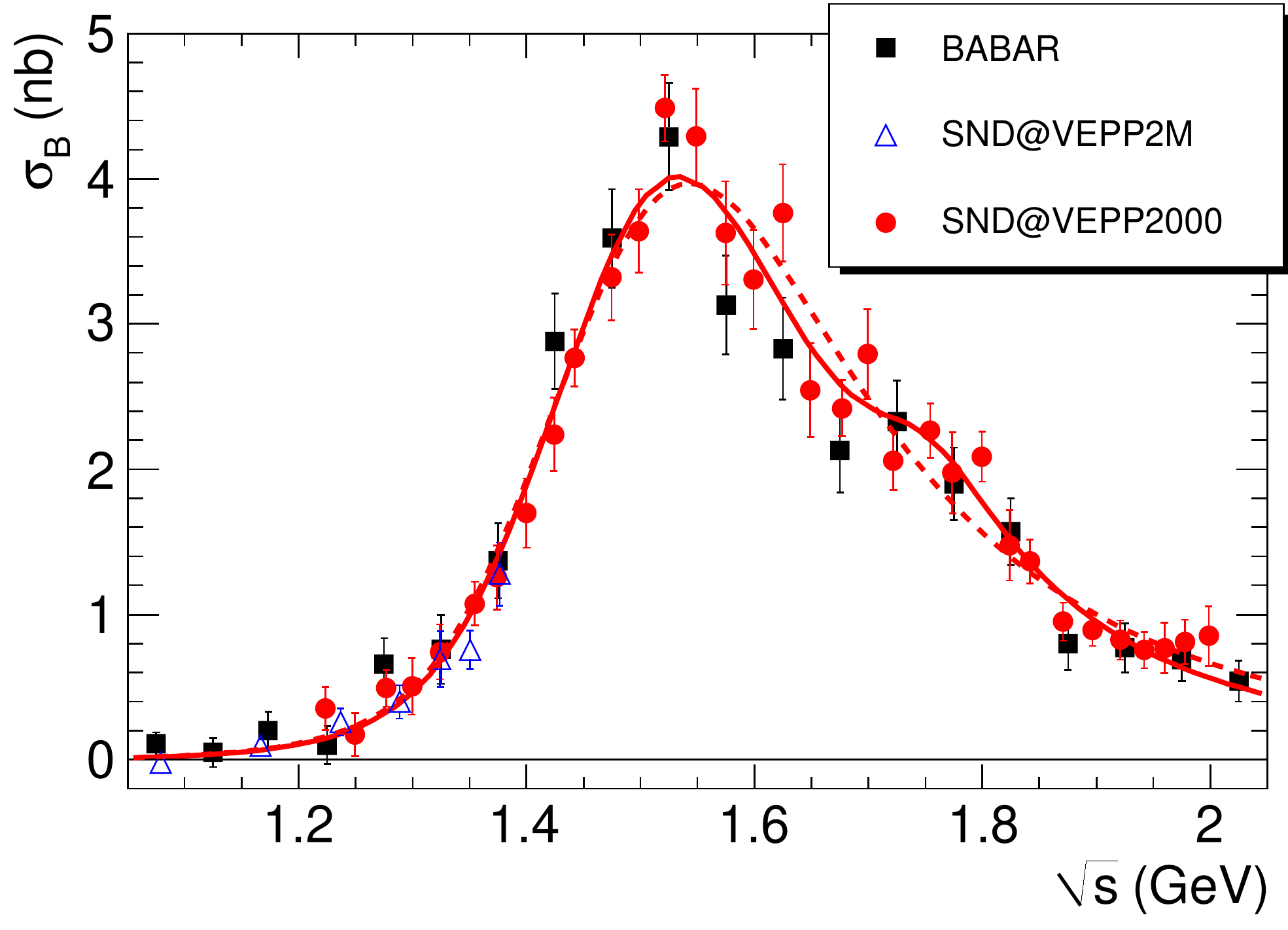}
\caption{The Born cross section for
$e^+e^- \to \eta \pi^+\pi^-$ measured in this (SND@VEPP2000) and previous 
experiments (BABAR~\cite{bib:babar} and SND@VEPP2M~\cite{bib:eppvepp2m}).
The solid curve is the result of the VMD fit with the $\rho(770)$,
$\rho(1450)$ and $\rho(1700)$ resonances. The dashed curve is the
same fit without the $\rho(1700)$ contribution.
\label{fig:bcrs}}
\end{figure}

The Born cross section for \epp 
obtained using Eq.~(\ref{eq:expbcrs}) is shown in Fig.~\ref{fig:bcrs}
in comparison with the results of the most precise previous measurements
by SND at VEPP-2M~\cite{bib:eppvepp2m} and BABAR~\cite{bib:babar}.
The numerical values are listed in Table~\ref{tab:bcrs}. The quoted errors 
on the cross section are statistical. The systematic uncertainty
is 8.3\% at $\sqrt{s}<1.45$ GeV, 5.0\% at $1.45<\sqrt{s}<1.60$ GeV,
and 7.8\% at $\sqrt{s}>1.60$ GeV. It consists of the systematic uncertainty 
in background subtraction (Sec.~\ref{sec:subbg}), the uncertainty on the
detection efficiency (Sec.~\ref{sec:eff}), the model uncertainty on the 
radiative correction, and the error on the integrated luminosity (2\%).
It is seen that the data of all three experiments are in agreement.

The measured \epp Born cross section is fitted using the VMD model with
the three isovector states $\rho(770)$, $\rho(1450)$, and $\rho(1700)$
decaying to $\eta\rho(770)$~\cite{bib:thepp}: 
\begin{eqnarray}
    \sigma_\mathrm{B}(s) &=& \int\limits_{4m_\pi^2}^{(\sqrt{s}-m_\eta)^2} 
    \frac{d\sigma}{dq^2}dq^2, \label{eq:born}\\ 
    \frac{d\sigma}{dq^2}(s,q^2) &=& \frac{4\alpha^2}{3}\frac{1}{s\sqrt{s}}
    \frac{\sqrt{q^2}\Gamma_\rho(q^2)P_\eta^3(s,q^2)}{(q^2-m_\rho^2)^2+
      (\sqrt{q^2}\Gamma_\rho(q^2))^2}|F(s)|^2, \nonumber \\
    P_\eta^2(s,q^2) &=& [(s-m_\eta^2-q^2)^2-4m_\eta^2q^2]/{4s}, \nonumber \\
    \Gamma_\rho(q^2) &=& \Gamma_\rho(m_\rho^2)\frac{m_\rho^2}{q^2}
    \left(\frac{p_\pi^2(q^2)}{p_\pi^2(m_\rho^2)}\right)^{\frac{3}{2}}, \nonumber \\
	p_\pi^2(q^2) &=& {q^2}/{4}-m_\pi^2, \nonumber 
    \end{eqnarray}
    where $q$ is the 4-momentum of the $\pi^+\pi^-$ system, and $F(s)$ is
    the transition form factor for the vertex $\gamma^\ast\to\eta\rho$: 
\begin{equation}
    F(s)= \sum_V\frac{m_V^2}{g_{V\gamma}}\frac{g_{V\rho\eta}}
    {s-m_V^2+i\sqrt{s}\Gamma_V(s)},\textrm{~~}V=\rho(770),\rho(1450),\rho(1700).
    \label{eq:F2}
\end{equation}
Here $g_{V\rho\eta}$ and $g_{V\gamma}$ are the coupling constants 
for the transitions $V\to\rho\eta$ and $V\to\gamma^\ast$, respectively.
It is convenient to use notation $g_{V\rho\eta}/g_{V\gamma}= g_Ve^{i\phi_V}$.

In the fit, the mass and width of the $\rho(770)$ resonance are fixed at their
nominal values~\cite{bib:pdg}. The phase $\phi_{\rho(770)}$ is set to 0.
The coupling constants $|g_{\rho\rho\eta}|$ and 
$|g_{\rho\gamma}|$ are calculated using data on the partial widths for the 
decays $\rho(770)\to e^+e^-$ and $\eta\gamma$~\cite{bib:pdg}:
\begin{eqnarray}
    g_{\rho\gamma}^2 &=& \frac{4\pi}{3}\alpha^2\frac{m_\rho}{\Gamma(\rho \to e^+e^-)}, \label{eq:g} \\
    g_{\rho\eta\gamma}^2 &=& \frac{24}{\alpha}m_\rho^3\frac{\Gamma(\rho \to \eta\gamma)}
    {(m_V^2-m_\eta^2)^3}, \nonumber \\
    g_{\rho\rho\eta}&=&g_{\rho\gamma} g_{\rho\eta\gamma} \nonumber.
\end{eqnarray}

For the $\rho(1450)$ and $\rho(1700)$ resonances, the masses and widths are 
also fixed at the nominal values~\cite{bib:pdg}, but are allowed to be varied 
within their errors. The ratios $g_{\rho(1450)}$ and $g_{\rho(1700)}$ are
free fit parameters. Since the coupling constants are not expected to have
sizable imaginary parts, the fit is performed assuming that the phases
$\phi_{\rho(1450)}$ and $\phi_{\rho(1700)}$ are equal to zero or $\pi$.
The best value of $\chi^2/\nu=37/31$ ($P(\chi^2)\approx 20\%$), 
where $\nu$ is the number of degrees of
freedom, is obtained for the phase combination
$\phi_{\rho(1450)}=\phi_{\rho(1700)}=\pi$.
The fitted ratios of the coupling constants are 
\begin{eqnarray}
    g_{\rho(1450)}&=&0.48^{+0.05}_{-0.06} \mbox{ GeV}^{-1}, \label{eq:rat} \\
    g_{\rho(1700)}&=&0.02^{+0.03}_{-0.01} \mbox{ GeV}^{-1}, \nonumber
\end{eqnarray}
The fit result is shown in Fig.~\ref{fig:bcrs}. The obtained value of
$g_{\rho(1700)}$ deviates from zero by only $2\sigma$. So, we cannot draw 
a definite conclusion that the $\rho(1700)$ contribution is needed for
data description. For comparison, we show in Fig.~\ref{fig:bcrs} the result
of the fit with $g_{\rho(1700)}=0$. 
The $\chi^2/\nu$ value for this fit is $42.6/32$ ($P(\chi^2)\approx 10\%$).
The value of $g_{\rho(1450)}$ is used to obtain the product of the branching
fractions
\begin{equation}
    B(\rho(1450)\to\eta\pi^+\pi^-)B(\rho(1450)\to e^+e^-) = 
    (4.3^{+1.1}_{-0.9}\pm 0.2)\times 10^{-7},
\end{equation}
where the second error is systematic. This result can be compared with 
the same products for other $\rho(1450)$ decays:
$B(\rho(1450)\to\omega\pi)B(\rho(1450)\to e^+e^-)=
(5.3\pm0.4)\times 10^{-6}$~\cite{bib:kardapo} and
$B(\rho(1450)\to\pi^+\pi^-)B(\rho(1450)\to e^+e^-)=
(5.6\pm1.8)\times 10^{-7}$. The later product is calculated using
the parameters of the VMD fit to the $e^+e^- \to \pi^+\pi^-$ cross section
performed in Ref.~\cite{bib:pipi1}. We obtain the following
ratios of the branching fractions
\begin{equation}
    B(\rho(1450)\to\omega\pi):B(\rho(1450)\to\eta\pi^+\pi^-):B(\rho(1450)\to\pi^+\pi^-)
    =12.3\pm3.1 : 1 : 1.3\pm0.4.
\end{equation}
There are several theoretical predictions for these ratios,  for example,
$8.1:1:9.5$~\cite{bib:isgur} and $6.4:1:3.8$~\cite{bib:donnachie}. 
It is seen that the experimental ratio 
$B(\rho(1450)\to\omega\pi)/B(\rho(1450)\to\eta\pi^+\pi^-)$ is in reasonable 
agreement with the predictions, while the $\rho(1450)\to\pi^+\pi^-$ decay 
rate  is too small compared to the theoretical expectations.

Under the CVC hypothesis, our data on the \epp cross section can be used to
calculate the branching fraction of the 
$\tau^- \to \eta\pi^-\pi^0\nu_\tau$ decay~\cite{bib:taubr} 
\begin{equation}
    \frac{B(\tau^- \to \eta\pi^-\pi^0\nu_\tau)}{B(\tau^- \to \nu_\tau e^-\bar{\nu}_e)}=
    \frac{3\cos^2\theta_c}{2\pi\alpha^2 m_\tau^8}
    \int\limits_0^{m_\tau^2} dq^2 q^2 (m_\tau^2-q^2)^2(m_\tau^2+2q^2)
    \sigma_{e^+e^-\to\eta\pi^+\pi^-}(q^2).
    \label{eq:taubr}
\end{equation}
Performing numerical integration of the measured cross section, we obtain 
the branching fraction 
\begin{equation}
    B(\tau^-\to\eta\pi^-\pi^0\nu_\tau) = (0.156 \pm 0.004 \pm 0.010)\%,
\end{equation}
which is in agreement with the world average experimental value
$(0.139 \pm 0.010)\%$~\cite{bib:pdg} and with the CVC result
$(0.153\pm0.018)\%$~\cite{bib:eidelman} obtained using earlier \epp data.

\section{Summary}
In this paper the cross section for \epp has been measured in the c.m. energy
range from 1.22 to 2.00 GeV. Our data are in agreement with previous
measurements and most precise in the energy region between 1.4 and 2.0 GeV.

We have studied internal structure of the $\eta\pi^+\pi^-$ final state.
It has been found that the $\rho(770)\eta$ intermediate state is dominant, but
does not fully describe the observed $\pi^+\pi^-$ invariant mass spectrum.

The measured cross section is well described by the VMD model with the
$\rho(770)$
and $\rho(1450)$ resonances. Adding the $\rho(1700)$ contribution improves the fit
quality, but is not necessary at the current level of statistics. From the fit
we have extracted the product of the branching fractions 
$B(\rho(1450)\to\eta\pi^+\pi^-)B(\rho(1450)\to e^+e^-)$ and compared it 
with the same products for $\rho(1450)\to\omega\pi^0$ and 
$\rho(1450)\to\pi^+\pi^-$ decays.

The branching fraction of $\tau^- \to \eta\pi^-\pi^0\nu_\tau$ decay
has been calculated from our cross-section data under the CVC hypothesis. 
The obtained $B(\tau^- \to \eta\pi^-\pi^0\nu_\tau)$ value is in agreement 
with the current experimental value and has comparable accuracy. 
The CVC hypothesis for the $\eta\pi\pi$ system works within the experimental
accuracy of about 10\%.

\section*{ACKNOWLEDGMENTS}
This work is partially supported  in the framework of the State                 
order of the Russian Ministry of Science and Education,                          
by RFBR grants No. 12-02-01250-a, 14-02-31375-mol\_a, 13-02-00375 and       
the Russian Federation Presidential Grant for Scientific Schools NSh-2479.2014.2.

\end{document}